\documentclass[aps,prb,twocolumn,showpacs,groupedaddress]{revtex4}

\usepackage{amssymb}

\usepackage{graphicx}

\usepackage{bm}

\begin{document}
\title{Fractional Quantum Hall Effect in SiGe/Si/SiGe Quantum Wells\\ in Weak Quantizing Magnetic Fields}
\author{V.~T. Dolgopolov, M.~Yu.\ Melnikov, and A.~A. Shashkin}
\affiliation{Institute of Solid State Physics, Chernogolovka, Moscow District 142432, Russia}
\author{S.-H. Huang and C.~W. Liu}
\affiliation{Department of Electrical Engineering and Graduate Institute of Electronics Engineering, National Taiwan University, Taipei 106, Taiwan, and\\ National Nano Device Laboratories, Hsinchu 300, Taiwan}
\author{S.~V. Kravchenko}
\affiliation{Physics Department, Northeastern University, Boston, Massachusetts 02115, USA}

\begin{abstract}
We have experimentally studied the fractional quantum Hall effect (FQHE) in SiGe/Si/SiGe quantum wells in relatively weak magnetic fields, where the Coulomb interaction between electrons exceeds the cyclotron splitting by a factor of a few XX. Minima of the longitudinal resistance have been observed corresponding to the quantum Hall effect of composite fermions with quantum numbers $p = 1$, 2, 3, and 4. Minima with $p = 3$ disappear in magnetic fields below 7 Tesla, which may be a consequence of the intersection or even merging of the quantum levels of the composite fermions with different orientations of the pseudo-spin, \textit{i.e.}, those belonging to different valleys. We have also observed minima of the longitudinal resistance at filling factors $\nu = 4/5$ and 4/11, which may be due to the formation of the second generation of the composite fermions.
\end{abstract}
\maketitle

\par\bigskip
The fractional quantum Hall effect (FQHE), discovered in 1982 \cite{tsui}, is no longer a topic of intensive research; rather, it has become a subject of reviews and textbooks \cite{u,u1,u2,u3}. In the FQHE, the Hall resistance, $\rho_{\text{xy}}$, of a two-dimensional electron system, placed in a magnetic field ($B$), perpendicular to its surface, exhibits plateaus, accompanied with minima in the longitudinal resistance, $\rho_{\text{xx}}$.  These minima correspond to the jumps of the chemical potential, occurring at filling factors, $\nu$, defined as the ratio of the carrier density, $n_{\text{s}}$, to the number of flux quanta per unit area
\begin{equation}
\nu=\frac{n_s 2\pi \hbar c}{eB}= \nu_{\text{QHE}} ,
\label{eq1}
\end{equation}
where $\nu_{\text{QHE}}$ is equal to an integer for the integer quantum Hall effect (IQHE) \cite{klitz} or in the simplest case to a fraction with an odd denominator for the FQHE.  The corresponding values of the Hall resistance are

\begin{equation}
\rho_{\text{xy}}^{\nu} = \frac{2\pi \hbar}{e^2\nu_{\text{QHE}}} .
\label{eq2}
\end{equation}
The origin of the FQHE arises from electron-electron interactions \cite{u,u1,u2,u3}.

A successful theoretical description of the FQHE is based on the concept of composite fermions \cite{jain}, where the FQHE is mapped onto the IQHE for composite particles, consisting of an electron and two (or, for the smallest fractions, even four) flux quanta in an effective magnetic field $B^*$
\begin{equation}
B^*= B-2\pi n_{\text{s}} \tilde \phi,
\label{eq3}
\end{equation}
where the case of a composite fermion with two flux quanta is considered (here $\tilde \phi =2 \hbar c/e$).  Filling factors for the IQHE for composite fermions are
\begin{equation}
p = \frac{2 \pi n_s \hbar c}{eB^*},
\label{eq4}
\end{equation}
while the corresponding original filling factors are
\begin{equation}
\nu^{\text{QHE}} = \frac{p}{2p\pm1}.
\label{eq5}
\end{equation}
Interactions between original particles enter the theory implicitly because a mean field approximation is used, assuming that the fluctuations of the density of the original particles are small.

The concept of composite fermions is well-documented experimentally.  At $\nu=1/2$, the effective magnetic field $B^*= 0$ (see Eq.~(\ref{eq3})), and the composite fermions behave like ordinary fermions in zero magnetic field \cite{halp}. This leads to a new scale in momentum space: a Fermi momentum of composite fermions that exceeds the momentum of the original particles by $\sqrt 2$. The existence of this scale has been verified in several experiments \cite{weiss,weiss1,focus}.

However, in all these experiments the Coulomb energy, $E_c \simeq e^2/\varepsilon l$ (here $\varepsilon$ is the dielectric constant and $l$ is the magnetic length), is smaller or comparable to the cyclotron splitting, $\hbar \omega_{\text{c}}$. This holds, in particular, for the experiments on relatively low-mobility ($\mu\sim 250,000$ cm$^2$/Vs) SiGe/Si/SiGe quantum wells \cite{tsui3}, in which the FQHE was studied at electron density $n_{\text{s}}=2.7\times10^{11}$ cm$^{-2}$ and in magnetic fields $11-40$~T.

There is a significant advantage in using ultra-clean SiGe/Si/SiGe structures having an order of magnitude higher electron mobility, as the magnetic fields required to observe the FQHE shift down by more than a factor of four. This results in the Coulomb energy becoming much greater than the cyclotron energy. We should mention that some minima in the longitudinal resistance, corresponding to the FQHE, have been previously observed in SiGe/Si/SiGe structures in weak magnetic fields in Ref.\cite{tsui2}.

In this Letter, we report our study of the FQHE in exceptionally high-mobility SiGe/Si/SiGe quantum wells, in which the minima at certain fractional filling factors ($\nu$ = 4/3, 2/3) could be observed at electron densities down to $n_{\text{s}} =2\times10^{10}$ cm$^{-2}$. Minima at $\nu=3/5$ and 3/7 disappeared below $n_{\text{s}}\simeq 7\times10^{10}$ cm$^{-2}$, although the surrounding minima (at $p=2$ and $p=4$) survived even at much lower electron densities.  This behavior might be explained by crossing or even merging of the quantum levels of composite fermions with different spin orientations \cite{jain1}; however, this would require an anomalously small Lande $g$-factor of composite fermions.  More likely, this is a consequence of the intersections of the quantum levels of composite fermions belonging to different valleys.

Samples used are SiGe/Si/SiGe quantum wells similar to those described in detail in Ref.\cite{melnikov}. The approximately 150 \AA\ wide silicon (001) quantum well is sandwiched between Si$_{0.8}$Ge$_{0.2}$ potential barriers. The upper layer of SiGe, 1500 \AA\ thick, was covered by a 10 \AA\ thick layer of silicon; and an approximately 2000 -- 3000~\AA\ thick layer of SiO was then deposited on top of the Si layer in a thermal evaporator, and an aluminum gate was deposited on top of SiO. The samples were patterned in Hall-bar shapes with the distance between the potential probes of 150~$\mu$m and width of 50~$\mu$m using standard photo-lithography. The long side of the Hall bar corresponded to the direction of current parallel to the [110] or [-110] crystallographic axes. Measurements were carried out in an Oxford TLM-400 dilution refrigerator at a temperature $T\approx0.03$~K. Magnetoresistance was measured with a standard four-terminal lock-in technique in a frequency range 1--10~Hz in the linear response regime (currents used were kept below 1~nA).

\begin{figure}
\scalebox{0.8}{\includegraphics[width=\columnwidth]{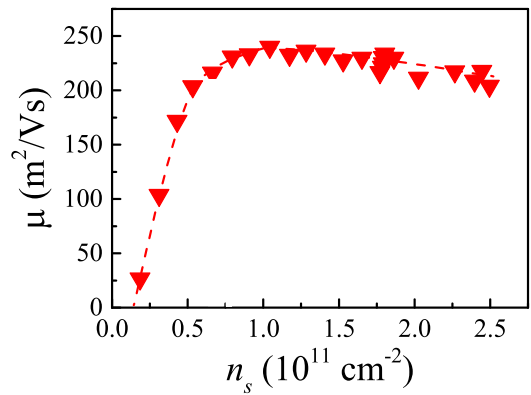}}
\caption{An example of the density dependence of the mobility in a SiGe/Si/SiGe quantum well at $T=50$~mK.}
\label{FigSiGe}
\end{figure}

The range of achievable electron densities in these samples was restricted from both above and below.  At high electron densities, the restriction was caused by the breakdown voltage of the dielectric, while at very low electron densities, the resistance of the contacts increased dramatically. The maximum electron mobility was about $240$~m$^2$/Vs at an electron density of $n_{\text{s}}\approx10^{11}$~cm$^{-2}$.  An example of density dependence of the mobility is shown in Fig.~\ref{FigSiGe}. The mobility varied from one cool-down to another and was increased by infra-red illumination.

\begin{figure}
\scalebox{1.0}{\includegraphics[width=\columnwidth]{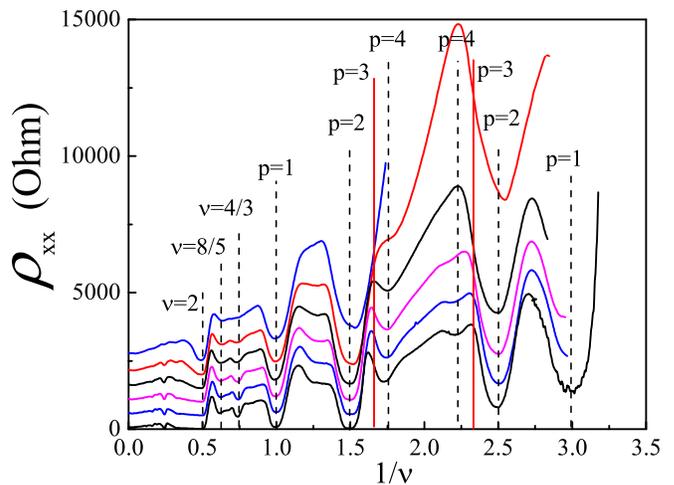}}
\caption{Longitudinal magnetoresistance of sample 1 at electron densities (from bottom to top): 6.33; 5.70; 5.08; 4.52; 3.85; and $3.05\times10^{10}$ cm$^{-2}$. Curves are vertically shifted by 500~Ohm for each step in density. Dashed vertical lines mark the experimentally observed minima of the resistance while solid vertical lines mark the minima that are expected, but not observed.}
\label{fig2}
\end{figure}

\begin{figure}
\scalebox{1.0}{\includegraphics[width=\columnwidth]{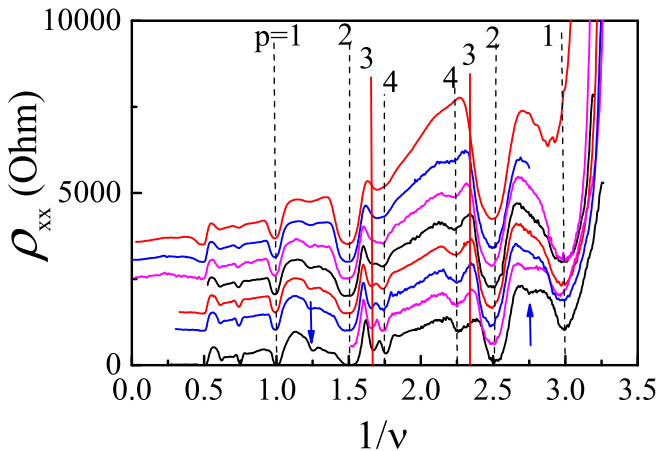}}
\caption{Longitudinal magnetoresistance of sample 2. Vertical lines are similar to those in Fig.\ref{fig2}, and the curves are also shifted by 500~Ohm. Electron densities (from bottom to top) are 9.31; 8.16; 7.56; 6.94; 6.31; 5.68; 5.10; and $4.43\times10^{10}$ cm$^{-2}$. Arrows designate $\nu=4/5$ and $\nu=4/11$.}
\label{fig3}
\end{figure}

The longitudinal magnetoresistance for two samples is shown in Figs.~\ref{fig2} and \ref{fig3}. In Fig.~\ref{fig2}, in the vicinity of $\nu=1/2$, minima of the resistance corresponding to two series of composite fermions (in positive and negative effective field $B^*$) are observed with $p=$~1, 2, and 4.  Minima at $p=3$ are absent at both $\nu>1/2$ and $\nu<1/2$.  The absence of the minima at $p=3$ cannot be due to the increasing width of the quantum levels of composite fermions because the minima at $p=4$ are clearly seen.

Sample 2 demonstrated a similar behavior.  In this sample, higher electron densities were reached, and the resistance minimum at $p=3$ developed (in negative $B^*$) at densities above $\sim6\times10^{10}$ cm$^{-2}$, becoming more pronounced with increasing density.  The minimum at $\nu=4/7$ is very narrow, which confirms that the absence of the minimum at $\nu=3/5$ cannot be due to the level broadening.

In Fig.~\ref{fig3}, two additional minima that are symmetric relative to $\nu=1/2$ are marked by arrows. One of them (at $\nu=4/5$) has been frequently observed earlier in high-mobility GaAs/AlGaAs heterostructures; the second one (at $\nu=4/11$) has been recently discovered in GaAs/AlGaAs \cite{tsui1}. Both of those minima correspond to the fractional filling factor of composite fermions, $p=4/3$, which suggests a formation of the second generation of composite fermions on the basis of already existing ones \cite{jain1}.

The absence of the resistance minima at $p=3$ at low electron densities (or, consequently, in weak magnetic fields) might be due to crossing or merging of the quantum levels of composite fermions with opposite spin orientations \cite{jain1,du,du1,QQ}. Indeed, in high magnetic fields, Zeeman energy, $\mu gB$, strongly exceeds the cyclotron energy of composite fermions that is proportional to $e^2/\varepsilon l$, and spin polarization of composite fermions at $p=3$ is the same for all three levels.  However, when the magnetic field is decreased, the lowest quantum level with the opposite spin orientation (originally empty), having an energy that is proportional to $B$, becomes coincident with the upper filled level, having an energy that is proportional to $B^{1/2}$.  Under these conditions, the energy gap and the corresponding resistance minimum at $p=3$ disappears.

When the magnetic field is further decreased, two scenarios are possible: either crossing of the quantum levels and reappearance of the gap; or merging of the levels so they become locked together \cite{dol,dol1}. In the first case, the spin polarization becomes independent of magnetic field and equal to 1/3 of the original one; in the second case, the spin polarization gradually changes from 1 to 1/3 with decreasing magnetic field.

According to Ref.\cite{QQ}, the energy gap at $p=3$ vanishes when the following condition is satisfied:
\begin{equation}
\mu g B \simeq 0.015 \frac{e^2}{\varepsilon l}.
\label{eqQQ}
\end{equation}

For electrons in a SiGe/Si/SiGe quantum well, in which $g=2$, the condition (\ref{eqQQ}) is met when $B_{\text{c}} \simeq 0.25$~T. This value dramatically differs from the experimentally measured $B_{\text{c}} \simeq 4.5$~T for $\nu=3/5$ and $B_{\text{c}} \agt 7$~T for $\nu=3/7$. To reach agreement between the results of Ref.\cite{QQ} and our experimental data on SiGe/Si/SiGe, one needs to assume that the $g$-factor in the latter case is smaller than the original one by a factor of at least four and is comparable to that in GaAs.

Another, more likely explanation of our results is crossing of the quantum levels of composite fermions belonging to different valleys.  The valley gap in a two-dimensional electron gas in Si decreases with decreasing electron density \cite{khrap} and, therefore, the above reasoning may well succeed for valley splitting.  Furthermore, in the latter case one may expect the quantum levels to be locked together similar to the results of Ref.\cite{dol}.

Finally, we would like to add that mixing of the quantum levels due to the Coulomb interactions did not manifest itself in our experiments.

We are grateful to I.~S. Burmistrov and D. Heiman for useful discussions.  ISSP group was supported by RFBR 18-02-00368, 16-02-00404 and RAS.  The support of NTU group by Ministry of Science and Technology, Taiwan, under the project numbers 106-2221-E-002-197-MY3, 106-2221-E-002-232-MY3, and 106-2622-8-002-001 is highly acknowledged.   SVK was supported by NSF Grant 1309337 and BSF Grant 2012210.

\par\bigskip
\par

\bigskip


\begin{thebibliography}{90}
\bibitem{tsui} D.~C. Tsui, H.~L. Stormer, and A.~C. Gossard, Phys.\ Rev.\ Lett.\ {\bf 48}, 1559 (1982).
\bibitem{u} T. Chakraborty and P. Pietilainen, The Fractional Quantum Hall Effect (Springer-Verlag, Berlin, New York, 1988).
\bibitem{u1} The Quantum Hall Effect, ed.\ by Richard E.\ Prange and Steven M.\ Girvin (Springer-Verlag, New York, 1990).
\bibitem{u2} Perspectives in Quantum Hall Effects, Edited by Sankar Das Sarma and Aron Pinczuk (Wiley , New York, 1997).
\bibitem{u3} S.~M. Girvin, `The Quantum Hall Effect: Novel Excitations and Broken Symmetries,' Les Houches Lecture Notes, in: Topological Aspects of Low Dimensional Systems, ed.\ by Alain Comtet, Thierry Jolicoeur, Stephane Ouvry, and Francois David, (Springer-Verlag, Berlin and Les Editions de Physique, Les Ulis, 2000).
\bibitem{klitz} K.\ von Klitzing, G.\ Dorda, and M.\ Pepper, Phys.\ Rev.\ Lett.\ {\bf  45}, 494 (1980).
\bibitem{jain} J.~K. Jain,  Phys.\ Rev.\ Lett.\ {\bf 63}, 199 (1989).
\bibitem{halp} B.~I. Halperin, P.~A. Lee, and N. Read, Phys.\ Rev.\ B {\bf47}, 7312 (1993).
\bibitem{weiss} J.~H. Smet, D. Weiss, K.\ von Klitzing, P.~T. Coleridge, Z.~W. Wasilewski, R. Bergmann, H. Schweizer, and A. Scherer, Phys.\ Rev.\  B {\bf 56}, 3598 (1997).
\bibitem{weiss1} J. H. Smet, S. Jobst, K. von Klitzing, D. Weiss, W. Wegscheider, and V. Umansky, Phys. Rev. Lett.,{\bf  83}, 2620 (1999)
\bibitem{focus} J. H. Smet, D. Weiss, R. H. Blick, G. Lutjering, K. von Klitzing, R. Fleischmann, R. Ketzmerick, T. Geisel, and G. Weimann, Phys. Rev. Lett. {\bf 77}, 2272 (1996).
\bibitem{tsui3} K. Lai, W. Pan, D.~C. Tsui, S. Lyon, M. Muhlberger, and F. Schaffler,  Phys.\ Rev.\ Lett.\ {\bf  93}, 156805 (2004).
\bibitem{tsui2} T.~M. Lu, D.~C. Tsui, C.-H. Lee, and C.~W. Liu, Appl.\ Phys.\ Lett.\ {\bf 94}, 182102 (2009).
\bibitem{jain1} K. Park and J.~K. Jain, Solid State Commun.\ {\bf 119}, 291 (2001).
\bibitem{melnikov} M.~Yu. Melnikov, A.~A. Shashkin, V.~T. Dolgopolov, S.-H. Huang, C.~W. Liu, and S.~V. Kravchenko, Appl.\ Phys.\ Lett.\ {\bf 106}, 092102 (2015).
\bibitem{tsui1} W. Pan, K.~W. Baldwin, K.~W. West, L.~N. Pfeiffer, and D.~C. Tsui, Phys.\ Rev.\ B  {\bf 91}, 041301 (2015).
\bibitem{du} R.~R. Du, A.~S. Yeh, H.~L. Stormer, D.~C. Tsui, L.~N. Pfeiffer, and K.~W. West, Phys.\ Rev.\ Lett.\ {\bf 75}, 3926
(1995).
\bibitem{du1} R.~R. Du, A.~S. Yeh, H.~L. Stormer, D.~C. Tsui, L.~N. Pfeiffer, and K.~W. West, Phys.\ Rev.\ B {\bf 55}, R7351 (1997).
\bibitem{QQ} I.~V. Kukushkin, K. von Klitzing, and K. Eberl, Phys.\ Rev.\ Lett.\ {\bf 82}, 3665 (1999).
\bibitem{dol} A.~A. Shashkin, V.~T. Dolgopolov, J.~W. Clark, V.~R. Shaginyan, M.~V. Zverev, and V.~A. Khodel,  Phys.\ Rev.\ Lett.\ {\bf 112}, 186402 (2014).
\bibitem{dol1} A.~A. Shashkin, V.~T. Dolgopolov, J.~W. Clark, V.~R. Shaginyan, M.~V. Zverev, and V.~A. Khodel, JETP Lett.\ {\bf 102}, 36 (2015).
\bibitem{khrap} V.~S. Khrapai, A.~A. Shashkin, and V.~T. Dolgopolov, Phys.\ Rev.\ B  {\bf 67}, 113305 (2003).
\end{thebibliography}
\end{document}